\documentstyle[12pt]{article}
\font\titolo=cmbx12 scaled\magstep2

\font\tsnorm=cmr12

\font\tscorsp=cmti10
\def\CQG{ Class. Quant. Grav. }

\def\NPB{Nucl. Phys. }
\def\PLB{ Phys. Lett.  }  

\def\PRD{Phys. Rev.  }

\def\CMP{ Commun. Math. Phys. }

\def\JHEP{ J. High Energy Phys. }
\def\z{Z\kern -4.6pt Z}
\def\dx{\int d^2x\ \sqrt{-g}\ }

\def\l{\lambda}

\def\a{\alpha}

\def\c{\chi}
\def\d{\delta}
\def\ep{\epsilon}
\def\f{\phi}
\def\g{\gamma}

\def\o{\omega}

\def\r{\rho}
\def\s{\sigma}

\def\y{\eta}
\def\z{\zeta}

\def\lie{{\cal L}}
\def\de{\partial}

\def\inf{\infty}
\def\id{\equiv}
\def\mo{{-1}}

\def\mn{{\mu\nu}}  

\def\ds{ds^2=}

\def\as{asymptotic symmetries }

\def\d{\delta}
\def\e{\eta}
\def\eo{\e_0}
\def\m{\mu}
\def\n{\nu}

\def\dx{\int d^2x\ \sqrt{-g}\ }
      
\def\ord#1{o\left(#1\right)}

\def\Pe{\P_\y}
\def\Ps{\P_\s}

\def\r{\rho}
\def\s{\sigma}
\def\x{\chi}
\def\ds{ds^2=}

\def\la{\l^2}

\def\be{\begin{equation}}
\def\ee{\end{equation}}
\def\bea{\begin{eqnarray}}
\def\eea{\end{eqnarray}}
\def\bc{\begin{displaymath}}
\def\ec{\end{displaymath}}
\def\lb{\label}

\def\adsd{$\rm AdS_{2}$ }

\def\adscft{$\rm AdS_{2}/CFT_{1}$ }
\def\gtt{\g_{tt}}
\def\gxx{\g_{xx}}

\def\gff{\g_{\f\f}}
\def\ab{asymptotic behaviour }
\def\diff{diffeomorphism }
\def\Pe{\Pi_\y}
\def\Ps{\Pi_\s}

\begin{document}
\pagestyle{empty}
\null
\vskip 5truemm
\begin{flushright}
INFNCA-TH0004\\
\end{flushright}
\vskip 15truemm
\begin{center}
{\titolo   Symmetry Breaking, Central Charges     }
\end{center}
\begin{center}
\titolo{ and the 
AdS$_{2}$/CFT$_{1}$ Correspondence }
\end{center}
\vskip 15truemm
\begin{center}
{\tsnorm Mariano Cadoni$^{a,c,*}$ and Salvatore Mignemi$^{b,c,**}$}
\end{center}
%\smallskip
\begin{center}
{$^a$\tscorsp Dipartimento di Fisica,  
Universit\`a  di Cagliari,}
\end{center}
%\smallskip
\begin{center}
{\tscorsp Cittadella Universitaria, 09042, Monserrato, Italy.}
\end{center}
%\smallskip
%\smallskip
\begin{center}
{\tscorsp $^b$  Dipartimento di Matematica, Universit\`a  di 
Cagliari,}
\end{center}
%\smallskip
\begin{center}
{\tscorsp viale Merello 92, 09123, Cagliari, Italy.}
\end{center}
%\smallskip
\begin{center}
{\tscorsp $^c$  INFN, Sezione di Cagliari.}
\end{center}
\vskip 19truemm
%\baselineskip=2\normalbaselineskip
%%%%%%%%%%%%%%%%%%%%%%%%%%%%%%%%%%%%%%%%%%%%%%%%%%%%%%%%%%%%%%%%%%%%%%
%%                    abstract                                 %%
%%%%%%%%%%%%%%%%%%%%%%%%%%%%%%%%%%%%%%%%%%%%%%%%%%%%%%%%%%%%%%%%%%%%%%
\begin{abstract}
\noindent
When two-dimensional Anti-de Sitter space (\adsd) is endowed with a 
non-constant 
dilaton the origin of the central charge in the Virasoro algebra 
generating the asymptotic symmetries of \adsd can be traced back to 
the breaking of the $SL(2,R)$ isometry group of \adsd.
We use this fact to clarify some controversial results appeared in the 
literature about the value of the central charge in these models. 
\end{abstract}
%%%%%%%%%%%%%%%%%%%%%%%%%%%%%%%%%%%%%%%%%%%%%%%%%%%%%%%%%%%%%%%%%%%%%%
%%                         End of abstract                          %%
%%%%%%%%%%%%%%%%%%%%%%%%%%%%%%%%%%%%%%%%%%%%%%%%%%%%%%%%%%%%%%%%%%%%%%
%%%%%%%%%%%%%%%%%%%%%%%%%%%%%%%%%%%%%%%%%%%%%%%%%%%%%%%%%%%%%%%%%%%%%%
%%                          Address                                 %%
%%%%%%%%%%%%%%%%%%%%%%%%%%%%%%%%%%%%%%%%%%%%%%%%%%%%%%%%%%%%%%%%%%%%%%
\vfill
%\begin{flushleft}
%{\tsnorm PACS: 04.70.Dy, 04.50.+h\hfill}
%\end{flushleft}
%\begin{flushleft}
%{\tsnorm KEYWORDS: two-dimensional gravity models, black holes\hfill}
%\end{flushleft}
%\smallskip
%\vfill
\hrule
\begin{flushleft}
{$^*$E-Mail: CADONI@CA.INFN.IT\hfill}
\end{flushleft}
\begin{flushleft}
{$^{**}$E-Mail: MIGNEMI@CA.INFN.IT\hfill}
\end{flushleft}
\eject
\pagenumbering{arabic}
\pagestyle{plain}

The Anti-de Sitter(AdS)/conformal field theory (CFT) correspondence 
in two spacetime dimensions \cite{g1, CM1, Na}  
seems to contradict the general belief  that 
low-dimensional physics is simpler than the higher-dimensional one.
Compared with the higher-dimensional cases, the \adscft duality 
has many puzzling and controversial features.
Even though general arguments suggest that gravity on \adsd should be 
related to some sort of conformal mechanics living on the 
one-dimensional boundary of \adsd, the realization of this correspondence 
is rather involved. 
Although it has been shown that, analogously to the 
three-dimensional (3D) case \cite{3d}, the conformal symmetry involved  
is infinite 
dimensional \cite{CM1},  the search for a physical system realizing this 
symmetry has not been very successful.
On the other hand, the fact that \adsd has a disconnected boundary makes 
the \adscft correspondence problematic even at a more fundamental 
level.

At a technical level there are features of gravity on \adsd
that represent a further difficulty for sheding light on the subject, 
e.g.  the existence of several two-dimensional (2D) gravity 
models admitting \adsd as solution ( models with or without
dilatons) or  the fact that the boundary of \adsd is one-dimensional.

In a previous work \cite{CM1} , we have been able, working in the context 
of 2D 
dilaton gravity, to show that the asymptotic symmetry group of \adsd 
is generated by a Virasoro algebra. Using a canonical realization of 
the symmetry we have also computed the central charge of the algebra. 
Unfortunately, using our value of the central charge for the 
computation of the statistical entropy of 2D black holes, we found  
a discrepancy of a factor of $\sqrt 2$  with respect to
the thermodynamical entropy.

Later Navarro-Salas and Navarro \cite{Na}, using a boundary field
realization of 
the symmetry,  found a value of the central charge (half of 
our result) that produces a statistical entropy in agreement with 
the  thermodynamical one.

In this letter we try to clarify this controversial point. We show
that in models where \adsd  is endowed with a 
non-constant dilaton, the origin of the central charge of the Virasoro 
algebra can be traced back to the breakdown of the $SL(2,R)$ isometry
group of \adsd.
The authors of Ref. \cite{Na} use instead a non-scalar dilaton, 
which enables 
them  to keep the $SL(2,R)$ symmetry of \adsd unbroken.
However, using a non-scalar dilaton, they have to give up the 
diffeomorphisms invariance of the 2D dilaton gravity theory.
We will show that  keeping the dilaton to transform as a scalar, hence 
considering 
a truly  diffeomorphisms invariance  2D dilaton gravity theory, 
our previous result of Ref. \cite{CM1} for the central
charge can be recovered also adopting a boundary field realization of 
the asymptotic symmetry 
of \adsd.

 Let us consider a generic two-dimensional (2D) dilaton gravity model,
\be\lb{e1}
S ={1\over 2}\dx\left[\e R +\l^2 V(\e) \right],
\ee
where $\e$ is a scalar
field related to the usual definition of the dilaton $\phi$ by
$\e=\exp(-2\phi)$.
In the discussion of the symmetries of the model 
a crucial role is played by the scalar $\e$.   
In general, the scalar character of $\e$ implies that the symmetries
(isometries) of the 2D spacetime are broken by a non-costant dilaton.
In fact under the isometric transformations generated by a Killing 
vector $\x$, 
\be\lb{e2}
\d\e=\lie_\x\e=\x^\m\de_\m\e.
\ee
A non-constant  dilaton in general implies $\lie_\x\e\neq 0$.
On the other hand, 2D dilaton gravity always admits a Killing vector 
given by \cite{mann}:
\be\lb{e3}
\x_{(1)}^{\m}=\epsilon^{\m\n}\de_{\n}\e,
\ee 
which leaves $\e$ invariant.
Thus, a non-constant field $\e$ in general breaks down the isometry group 
of the metric to the subgroup generated by the Killing vector 
$\x_{(1)}$.  

The previous features of 2D dilaton gravity theories have strong 
analogies  with spontaneous  symmetry breaking in ordinary field theory.
In the case under consideration the quantity characterizing the breaking 
of the symmetry is  $\de_{\m}\e$.
The analogy is particularly evident if one considers classical solutions
of the model (\ref {e1}). These solutions are characterized by a mass $M$ 
and a temperature $T$ (when the solutions can be interpreted as black 
holes). Both $M$ and $T$ are geometric objects, so that they are 
invariant under the isometry group of  the metric, and  can be 
expressed in terms of $\e$ \cite{mann, cadoni}
\be\lb{e4}
M=F_{0}\left[\int d\e \la V(\e)- (\nabla\e)^{2}\right],\qquad\qquad
T\propto
V(\e_{h}),
\ee
where $F_{0}$ is a constant related to the normalization of the 
Killing vectors (in the following we use $F_{0}={1/2\l\eo})$ and
$\e_{h}$ is the scalar evaluated on the horizon of the 2D black 
hole. Using the field equations one can easily show that the 
solutions  characterized by a constant dilaton, thus preserving the 
isometry group of the metric, have zero mass and 
temperature. Only  symmetry-breaking  excitations, 
 characterized  by a non 
constant dilaton,  can have $M\neq 0$, $T\neq 0$. 

Let us now apply the previous  considerations to \adsd.  
With a dilaton potential $V=2\e$ the action (\ref{e1}) 
describes the Jackiw-Teitelboim (JT) model \cite{JT}. 
\adsd or more generally black holes in 
\adsd, are solutions of the model \cite{CM}. Owing to Birkhoff's 
theorem, the solutions can always be written in a static 
form, with dilaton $\e=\eo \l x$, where $\eo$ is an integration constant
(the field equations, but not the action, are invariant for rescaling of
$\y$ by a constant) and metric
\be\lb{e5}
\ds-(\l^2x^2-a^2)dt^2+(\l^2x^2-a^2)^{-1}dx^2,
\ee
where $a^2$ is proportional to the mass of the black hole.

Being a maximally symmetric
spacetime \adsd admits three Killing vectors generating the   
$SO(1,2)\sim SL(2,R)$ group of isometries. It is evident that the 
 static solution for $\e$ is invariant only under the action 
of the Killing vector (\ref{e3}), which in this case describes
time-translations $T$. Thus, the non-constant value of the dilaton   
breaks $SL(2,R)\to T$.  The parameter characterizing the symmetry 
breaking  is $\de_{x}\e=\eo \l$.

Similar considerations hold when one considers the \as of \adsd.
These are defined as the transformations which leave the asymptotic form
of the \adsd metric invariant, and where shown in \cite{CM1} to generate
a Virasoro algebra with non-trivial central charge.
Our aim here is to give a realization of this algebra in terms of fields
which describe the degrees of freedom of the boundary. For this purpose,
it is useful to adopt the formalism introduced in \cite{Na}.

We define  a two-dimensional metric to be asymptotically \adsd if, for
$x\to\inf$, it behaves as 
\bea
g_{tt}&=&
-{\l^2x^2 }+\gamma_{tt}(t)+\ord{1 \over x^2} \, ,\nonumber \\
g_{tx} &=& {\gamma_{tx}(t) \over \l^{3}x^3}+\ord{1 \over x^5} \, ,
\nonumber\\
g_{xx} &=& {1\over \l^2  x^2}+{\gamma_{xx} (t)\over \l^4 x^4}+
\ord{1\over x^6} \, ,
\label{d1}
\eea
where the fields $\g_\mn$ parametrize the first sub-leading terms in the
expansion and can be interpreted as deformations on the boundary.

The asymptotic form (\ref{d1}) of the metric is preserved by infinitesimal
diffeomorphisms $\c^\m(x,t)$ of the form \cite{CM1}
\bea
\c^t &=& \epsilon (t)+{\ddot\ep(t)\over 2\l^4x^2}+
{\alpha^t(t) \over x^4}+\ord{1 \over x^5} \, , \nonumber \\
\c^x &=& -x\dot\epsilon(t)+{\alpha^x(t) \over x}+\ord{1 \over x^2} \, .
\label{d2}
\eea
where $\ep(t)$ and $\alpha^\nu(t)$ are arbitrary and a dot denotes  
time derivative. $\alpha^\nu$ describes 
 "pure gauge"  2D diffeomorphisms which affect only the fields
on the boundary.
In Ref. \cite{CM1} is shown that the symmetries (\ref{d2}) are generated by
a Virasoro algebra. 

In view of (\ref{e2}), the \ab of the scalar field $\y$, compatible
with the transformations (\ref{d2}), must take the form
\be
\y=\eo\left(\l\r(t)\, x+{\gff(t)\over 2\l x}\right)+\ord{1 \over x^3},
\label{d3}
\ee
where $\r$ and $\gff$ play a role analogous to that of the $\g_\mn$.
On shell hold the constraints
\be
\l^{-2}\ddot\r=\r(\gtt-\gxx)-\gff ,\label{d4}
\ee
\be
\dot\r\gtt+{\r\over2}\dot\gxx+\dot\gff=0 .\label{d4a} 
\ee

Eq. (\ref{d3}) implies that the full, infinite 
dimensional, asymptotic symmetry of \adsd is broken by the boundary 
condition for $\e$. In fact, only the asymptotic transformations 
generated by the
Killing vector (\ref{e3}), which correspond to $\ep=\r$ in Eqs. 
(\ref{d2}),  leave $\e$ asymptotically invariant. The quantity 
characterizing the 
symmetry breaking is now the boundary field $\r=\de_x\e+o(x^{-1})$.

In \cite{Na} it was assumed that $\r(t)\id 1$. However, this is at variance
with the transformation (\ref{e2}) of $\y$ under diffeomorphisms, which
implies that the background dilaton $\y=\y_0\l x$, is transformed by
(\ref{d2}) into the form (\ref{d3}) with generic $\r(t)$.
In order to avoid this problem,
the authors of \cite{Na} had instead to assume that under diffeomorphisms
the dilaton transform as
$\d\y=\c^\m\de_\m\y+\de_t\c^t\y$ \cite{Napr}, but this seems rather ad hoc
and moreover spoils the \diff invariance of the action (\ref{e1}), 
presumably
leading to inconsistencies. It is important to notice that, if $\y$
transforms according to the previous
transformation law, the asymptotic symmetry 
group of \adsd is no longer broken.
In fact, in this case, asymptotically $\d\e=0$
under the full group generated by the Killing vectors (\ref{d2}). 

Although the  asymptotic symmetry of \adsd is broken by  
$\e$, the boundary fields $\gtt$, $\gxx$, $\gff$, $\r$
still span a representation of the full infinite dimensional group 
generated by the Killing vectors (\ref{d2}).   
In fact, under the \as (\ref{d2}), the boundary fields transform as
\bea
\d \gtt &=& \ep\dot\gtt+2\dot\ep\gtt-{\stackrel\dots\ep\over \l^{2}}
-2\la\a^x,
\nonumber  \\
\d \gxx &=& \ep\dot\gxx+2\dot\ep\gxx -4\la \a^x,\nonumber   \\
\d \gff &=& \ep\dot\gff+\dot\ep\gff+{\ddot\ep\dot\r\over\l^{2}}
+2\la\r\a^x ,
\nonumber \\
\d \r &=& \ep\dot\r-\dot\ep\r .                       
\label{d5}
\eea

These transformations are easily recognized as (anomalous) 
transformation laws for conformal fields of weight respectively 
$2,2,1,-1$. The anomalous parts of the transformation (the terms 
proportional to $\ddot \ep$ and $\stackrel\dots\ep$)  are related with 
the $SL(2,R)$ symmetry breaking.
In fact, the anomalous terms are connected one with the other by using the 
equation of motion (\ref{d4}), whereas the anomalous term appearing 
in the transformation of $\gff$  is proportional to $\dot \r$. 
It follows that, when \adsd 
is endowed with a non-constant dilaton, the origin of the anomalous 
terms in the 
transformation laws for conformal boundary fields (hence of the central 
charge) 
can be traced back to the  breakdown of the full asymptotic isometry 
group of the spacetime. This implies that in general one has a 
$\r$-dependent central charge.  However, we are only interested in 
classical solutions  of the 2D gravity model.  Because of Birkhoff's 
theorem, we can, without loss of generality, limit ourselves to the 
field configurations with $\r=$ const. Moreover, the equations of 
motions of the JT model are invariant under the rescaling of $\e$ by 
a constant. It will therefore be sufficient to consider only the $\r=1$ 
configuration.

The next step in our analysis is the construction of the generator of the 
conformal symmetry  in terms of the boundary fields 
$\gtt, \gxx, \gff, \r$. This generator has a natural interpretation as 
the stress-energy tensor associated with the one-dimensional conformal 
field theory living on the boundary of \adsd. A natural candidate for 
such a generator is the charge $J(\ep)$, which in the canonical 
formalism can be associated with the asymptotic symmetries  (\ref{d2}). 
The charge $J(\ep)$ is defined in terms of the boundary contribution
one must add to the Hamiltonian in order to have well defined variational 
derivatives \cite{CM1}, 
\be
\d J = - \lim_{x\to\inf}[N(\s^\mo\d\y'-\s^{-2}\y'\d\s)-N'(\s^\mo\d\y)
+N^x(\Pe\d\y-\s\d\Ps)]. \lb{d10}
\ee

Using the boundary conditions (\ref{d1}), (\ref{d3}) one finds,

\be\lb{d11}
\d J[\ep]=\eo\left[\l\ep(\gtt\d\r+{\r\over2}\d\gxx+\d\gff)+
{\dot\ep\d\dot\r\over\l}-{\ddot\ep\d\r\over \l}\right]\, .
\ee

This expression is locally integrable in a neighborhood of the classical
solutions \cite{CM1},
but unfortunately is not integrable globally in the full phase space.
The non-integrability of $\d J$ is, again, a consequence of the 
$SL(2,R)$ symmetry breaking. For instance, if one uses the 
symmetry-preserving 
boundary conditions for $\e$ used in Ref. \cite{Na}, $\d J$ is 
globally integrable and Eq. (\ref{d11}) yields the form of $J$ 
obtained in that paper.
Fortunately, we do not need  to know $J$ globally, but is sufficient to
know the form of $J$ near the $\r=1$ configuration. 
The expression of $J$ can be further simplified considering  
only on-shell field configurations.
Using the equations of motion (\ref{d4}) and (\ref{d4a}), and expanding
around the classical solutions, $\r=1+\bar\r$ in Eq. (\ref{d11}), 
one obtains
at leading order in $\bar\r$,

\be \lb{d13}
J(\ep)= {\eo\over 
\l}\left(\dot\ep\dot{\bar\r}-\ddot\ep\bar\r\right)+\ep M,
\ee
Where $M$ is the mass (\ref{e4}), which on shell becomes constant.
The charge (\ref{d13}) is defined up to an additive constant, 
we use a normalization such that $J(\ep=1)=M$. In the following we 
will consider,  for sake of 
semplicity, only the case $M=0$, i.e variations near the ground state. 
We are mainly interested in the value of the central charge of the 
Virasoro algebra, which is independent of $M$.
 
For generic $\ep$ the charges $J(\ep)$ are not conserved. The only 
conserved charge is obtained for $\ep=1$.
This fact is a consequence of the  $SL(2,R)\to T$ symmetry breaking:
only the charge associated with the residual symmetry $T$ is conserved.
This fact makes it impossible to use the 
charges $J(\ep)$ to give a realization of the Virasoro algebra, as 
it has been already shown in the canonical framework
\cite{CM1}. To solve the problem, we proposed to introduce the 
time-integrated charges,
\be\lb{d14}
\hat J(\ep)={\l\over 2\pi}\int_{0}^{2\pi\over \l}dt \,J(\ep).
\ee
The charges $\hat J$ are now trivially conserved and generate the 
asymptotic symmetries of \adsd trough the relation \cite{CM1}
\be\lb{d15}
\widehat{ \d_{\o}J(\ep)}=\left[ \hat J(\ep),\hat J(\o) \right],
\ee
where the hat means overall time-integration as defined in Eq. 
(\ref{d14}).

The definitions (\ref{d14}) imply that $J$ is defined up to a 
total time-derivative. Using this freedom, we can always write, 
\be\lb{f1}
J(\ep)= -2{\eo\over \l}\ep\ddot{\bar\r}=\ep \Theta_{tt}.
\ee
Using the transformation laws (\ref{d5}), one gets

\be\lb{f2}
\ep \d_{\o}\Theta_{tt}=\ep\left(\o\dot\Theta_{tt}+2\dot\o
\Theta_{tt}\right)+
c(\ep,\o),
\ee
where $c(\ep,\o)={\eo\over\l}(\ddot\ep\,\dot\o-\ddot\o\,\dot\ep)$.

The previous equation tells us that $\Theta_{tt}$ can be considered 
as the one-dimensional stress-energy tensor associated with the 
conformal symmetry, whereas $\hat J(\ep)$ are the charges generating 
a central extension of the Virasoro algebra. Notice that the central 
charge $c$ in general is $\r$-dependent, the expression quoted above being
the central charge evaluated near $\r=1$.
Expanding in Fourier modes, 
\be
\hat J(\ep)= \sum_{m}a_{m}L_{m}, \qquad \ep= \sum_{m}a_{m}e^{i\l m t},
\qquad \Theta_{tt}=\sum_{m}L_{m}e^{-i\l m t},
\ee
and using  Eqs. (\ref{d15}) and (\ref{f2}) one finds that the
$L_{m}$ span a   Virasoro algebra,

\be\lb{f3}
\left[ L_{m},  L_{n}\right]=(m-n) L_{m+n}+{C\over 
12}m^{3}\d_{m+n},
\ee
with central charge $C=24\eo$.
The value of the central charge of the algebra is exactly the same as 
that found by us using a canonical realization of the symmetry 
\cite{CM1} and differs by a factor $1/2$ from that found by 
Navarro-Salas and Navarro \cite{Na}. The origin of the mismatch is 
easily understood. As already pointed out,   the authors of Ref. \cite{Na} 
use   different  boundary conditions 
and a different transformation law 
for the field  $\e$. They use a non-scalar dilaton and  $\r=1$, identically,
 so that the 
$SL(2,R)$ isometry of 
\adsd is not broken, and the 
origin of the central charge is completely different from our case.
In Ref. \cite{Na} the central charge arises as consequence of the 
anomalous term in the transformation law  of $\gtt$ (see Eq. (\ref{d5})).
This is very similar to the three-dimensional case \cite{3d}. 
Their results hold  only for a  2D dilaton gravity model that is not 
invariant under space-time diffeomorphisms.
Therefore, the mismatch between statistical and 
thermodynamical entropy of 2D black hole discussed in our previous work 
\cite{CM1} still remains an open question.

\end{document}